# Implementation and Evaluation of GBDI Memory Compression Algorithm Using C/C++ on a Broader Range of Workloads

Adeyemi Aina
Computer Science
Virginia Tech
Blacksburg VA USA
ainababs0@vt.edu

## ABSTRACT

Memory compression is an important approach in computer architecture for decreasing memory footprint and improving system performance. In this paper, we use C/C++ to develop a current memory compression algorithm; the Global Bases Delta Immediate (GBDI) algorithm, which was proposed at HPCA'2022. By using global bases and enabling deltas within the same block to vary in size, the GBDI compression algorithm decreases the size of encoded data. The goal of this research is to assess the effectiveness of the GBDI algorithm by examining its compression ratios under a broader range of workloads. Our research leads to a better knowledge of the GBDI algorithm's effectiveness and the potential benefits of memory compression techniques for various sorts of applications. Furthermore, our C/C++ version of the algorithm gives academics and practitioners a high degree of freedom over customizing the algorithm for individual workloads and optimizing its performance.

## I. INTRODUCTION

Due to the growing use of memory-intensive applications and large data sets, the reduction of memory utilization has emerged as a critical challenge in the field of computer architecture. Memory is expensive and uses a lot of energy, so memory compression techniques have been developed to improve system performance. The GBDI algorithm, which was suggested in the HPCA'2022 paper, is a promising algorithm in this field. To fully realize its potential, the GBDI algorithm's performance must first be assessed using a wider variety of workloads. This study suggests using C/C++ to implement the GBDI algorithm and a variety of benchmarks that represent various workload types to test its compression ratios.

### 1.1 Memory compression techniques

Memory compression techniques such as Huffman coding, Lempel-Ziv (LZ) compression, Gzip, and LZ4 are proposed in the literature, each with their own set of advantages and disadvantages. Huffman coding uses shorter codes for more frequent symbols and longer codes for less frequent symbols, whereas LZ compression is useful for compressing repetitive symbol sequences. Gzip and LZ4 are fast and effective, but they may not be as efficient in terms of compression ratios as other techniques. The compression technique used is determined by the specific application requirements and trade-offs. In addition to these techniques, Base Delta Immediate (BDI) compression assigns a base within each block of data and compresses the deltas between the values and their assigned bases, allowing the deltas within the same block to have a fixed size, which may be less effective for certain types of data.

## II. OVERVIEW OF THE GBDI COMPRESSION

### A. Baseline GBDI

GBDI compression is a technique that reduces the size of encoded data by selecting several global bases across all targeted data. It differs from Block Delta Immediate (BDI) compression in that each block is assigned a base(Angerd et al., 2022). GBDI allows the size of deltas within the same block to vary by pairing each global base with a maximum delta. The algorithm has five steps, including calculating the delta for each value and determining the closest global base for each value. To determine global base values, GBDI employs modified Kmeans clustering. This technique outperforms BDI and achieves higher compression ratios than unmodified Kmeans.

### B. Implementing the GBDI Algorithm using C

The Global Bases Delta Immediate (GBDI) compression algorithm is a lossless data compression technique that can achieve high compression ratios for various workloads by identifying and encoding repeating patterns within input data. Because of its low-level nature, implementing the GBDI algorithm in C/C++ can be beneficial, allowing for efficient memory usage and faster execution times (Angerd et al., 2022). To implement the GBDI algorithm in C/C++, deep understanding of the algorithm's basic principles.

1) *Establishing Global Base Values:*

GBDI uses global bases, shared by all input data, instead of selection of intra-block (per block) base value(s) as in BDI.
through background data analysis, the global bases are established with Kmeans Clustering. The goal of the Kmeans Clustering is to find a set of "global bases" that can represent the data efficient. The kmeans algorithm goes through the following stages. 1) initializes k clusters centers randomly from the memory block. 2) Assign each data point of the memory block to the nearest cluster center. 3) then the Kmeans clustering algorithm updates the luster centers by

**Implementational and Evaluation of GBDI Memory Compression Algorithm Using C/C++ on a Broader Range of Workloads**

calculating the mean of all the data points of the memory block assigned to the cluster. 4) then steps 2 and 3 until convergence or it reaches a convergence threshold.

Once the clustering is complete, the centroids represent the global bases. Each data value can then be replaced with a pointer to its nearest global base (centroid) and a delta value representing the difference between the original value and the global base.

2) *Establishing Maximum Deltas:*

The global base is then subtracted from each element within the group, generating a series of deltas. The deltas determine the maximum allowable difference between consecutive data elements within the group. Take a global base B0, following the completion of the background data analysis then: the difference of the global base and values of the uncompressed blocks $B0 - V0 = \Delta0\ldots$

3) Encoding

The next step is to apply the encoding scheme to the parameters, the deltas, outliers, and base pointers are packed and encoded as well.

## III. RELATED WORK

The review analyzes several papers proposing solutions for memory compression and evaluates their effectiveness using synthetic and real-world workloads. The approaches discussed in the papers include rank subsetting, dynamic management of resources, novel clustering algorithms, and adaptive policies.

MemZip is a compressed memory architecture that uses rank subsetting to increase memory capacity, reduce page fault rates, and decrease energy and bandwidth demands. MemZip achieves a 45% performance improvement and a 57% reduction in memory energy compared to an uncompressed non-sub ranked baseline (Tsai & Sanchez, 2019).

A dynamic memory compression solution manages resources based on varying demands and situational requirements. It has demonstrated success under complex workload conditions and memory pressure, resulting in an increase in performance by up to 55 times across various benchmarks and applications depending on the size of the compressed area, the application's compression ratio, and access pattern used by the application.(Tuduce, 2005)

Global Base Delta Immediate (GBDI) is a novel compression method that offers substantially higher memory bandwidth than prior art. GBDI uses a clustering algorithm through data analysis to achieve this result. Using GBDI yields 1.5x higher bandwidth and 1.1x higher performance compared to a baseline without compression support on average when medium-high memory is required.

An adaptive policy for cache compression is proposed, which dynamically monitors a given workload and adjusts accordingly to achieve comparable benefits from compression while avoiding any unnecessary decompression overhead or degradation in overall performance.

## IV. IMPLEMENTATION

A. Compression Engine:

The compression engine reduces the size of the input data. The engine achieved through various algorithms and techniques that identify and eliminate redundancies within the data. The main components of a compression engine include:

1. Data analysis: the algorithm implements a statical algorithm to for efficient selection of the global bases. In the context of this paper, the Kmeans algorithm was implemented for the selection of Global bases.
2. Encoding: After selection of global bases and maximum deltas the compression engine converts the input data does the variable length encoding.

B. Decompression Engine

The decompression engine is responsible for reconstructing the original data from the compressed representation. This involves the following sub-processes:

*a. Format decoding:* The compressed data is first parsed to extract the necessary metadata, such as encoding information, compression algorithm details, and any other information required for decompression. This stage involves understanding the format of the compressed data and accurately interpreting its structure.

*b. Global table access*: The decompression engine may need to access global bases tables and maximum deltas array to aid in the reconstruction of the original data.

*c. Value* reconstruction: The decompression engine then reconstructs the original data by reversing the encoding process applied during compression. This involves decoding compressed data values, expanding patterns, or reconstituting sequences based on the information stored in the global tables. The accuracy of the value reconstruction process is essential to ensure that the decompressed data is identical or near identical to the original input.

## V EXPERIMENTAL METHODOLOGY

To evaluate the performance of the GBDI compression algorithm, the systematic experimental methodology approach to accurately assess the algorithm's effectiveness, efficiency, and overall impact was following.

*Data Selection*: We applied a diverse set of memory dump files in the ELF format, sourced from the CRC server. These files represent different types of workloads and applications, providing a comprehensive benchmark for evaluating the compression algorithm. The memory dump files include SPEC CPU 2017 benchmarks: 605.mcf_s_5.dump, 600.perlbench_s_5.dump, 620.omnetpp_s_5.dump,631.deepsjeng_s_5.dump.PARSEC benchmarks: parsec_fluidanimate5dump, parsec_freqmine5dump. JavaWorkloads: TriangleCount_3.dump, SVM_3.dump MatrixFactorization_4.dump

**Implementational and Evaluation of GBDI Memory Compression Algorithm Using C/C++ on a Broader Range of Workloads**

*Compression* We applied the compression algorithm to each of the selected memory dump files. Implementing the compression engine, which includes the global base delta algorithm, format encoding, and any additional optimizations, then applying the compression algorithm to each memory dump file, recording the compression ratio.

*Decompression:* After compressing the memory dump files, we performed the decompression process. This involves: Implementing the decompression engine, which includes format decoding, global table access, value reconstruction, and any additional optimizations. Decompressing each compressed memory dump file, recording the decompression time, and verifying the accuracy and correctness of the reconstructed data

*Performance Evaluation:* We evaluated the performance of the compression algorithm using various metrics, including:

- Compression ratio: The ratio of the original memory dump file size to the compressed file size.
- Reconstruction accuracy: The correctness and fidelity of the decompressed data when compared to the original memory dump files.

VI. RESULTS

These represents the expected outcome of the project:
1. Deep understanding of the GBDI algorithm and related memory compression techniques.
2. Implementation of the GBDI algorithm using C/C++ and validation of the implementation using a set of test cases.
3. Selection of benchmarks that represent different types of workloads.
4. Evaluation of compression ratios achieved by the GBDI algorithm using a subset of the selected benchmarks.
5. Preliminary conclusions about the effectiveness of the GBDI algorithm in a subset of the selected benchmarks.

The literature claimed a compression ratio of 1.9x (with K-means clustering). Implementing the compression algorithm, on average, the combination performs 1.55x for the Java workloads and 1.4x for the C-Workloads. Generally, the algorithm offers an average compression ratio of 1.4x

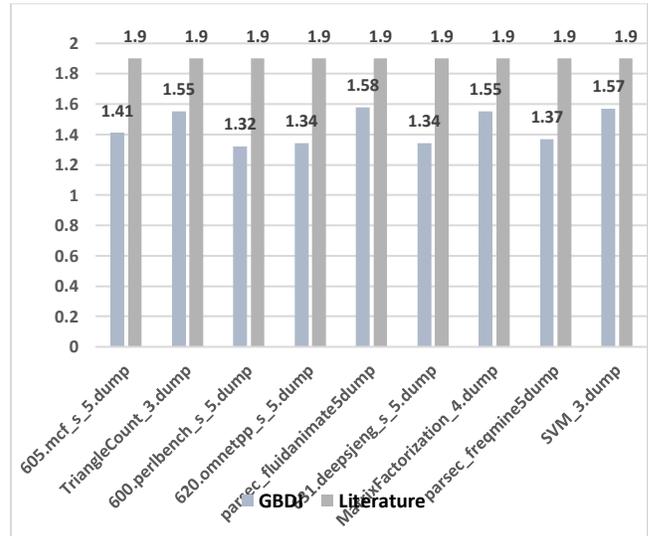

VII CONCLUSIONS

Global bases are established through background data analysis (Kmeans clustering) to find a base value that minimizes the distance to all cluster values, then further establishing maximum deltas in the Global Base Delta algorithm is a critical step in optimizing compression efficiency. Global-Base-Delta-Immediate Compression (GBDI) that exploits inter-block locality. The GBDI offers a compression ratio of an average of 1.45X with the Workloads.